\documentclass[twocolumn,aps,prl,showpacs,superscriptaddress,groupedaddress]{revtex4}
\usepackage{textcomp}
\usepackage{amssymb}
\usepackage{graphicx}

\begin{document}

\title{Cascading Quantum Light-Matter Interfaces}
\author{Mehdi Namazi, Thomas Mittiga, Connor Kupchak and Eden Figueroa}
\affiliation{Department of Physics and Astronomy, Stony Brook University, New
York 11794-3800, USA}

\begin{abstract}
The ability to interface multiple optical quantum devices is a key
milestone towards the development of future quantum networks that
are capable of sharing and processing quantum information encoded
in light. One of the requirements for any node of these quantum networks
will be cascadability, i.e. the ability to drive the input of a node
using the output of another node. Here, we report the cascading of
quantum light-matter interfaces by storing few-photon level pulses
of light in warm vapor followed by the subsequent storage of the
retrieved field onto a second ensemble. We demonstrate that even after
the sequential storage, the final signal-to-background ratio can remain
greater than 1 for weak pulses containing 8 input photons on average.
\end{abstract}


\maketitle

Any machine can be defined as a device composed of many constituents
with their own specific functions but when interfaced together, are
designed to carry out a much greater task. This same description would
hold true for a quantum information processor, a complex machine capable
of operating and processing quantum entities encoded with information.
Given the recent success in the creation and control of individual
quantum systems with a variety of physical architectures \cite{Reiserer2014,Northup2014},
the next logical step towards the realization of such a quantum machine
is the interconnection between multiple quantum interfaces \cite{Choi2010,Ritter2012,Vittorini2014,Duan2010}.
This type of functionality will be a prerequisite for networks in
which quantum information and entanglement can be shared, either sequentially
or simultaneously \cite{Perseguers2013,Krauter2013,Nunn2013}.

The success of these networks will rely on having universal quantum nodes producing outputs suited for driving (as inputs) succeeding quantum
nodes. This is the concept of quantum cascadability \cite{Miller2010}, and it is a necessary attribute for quantum computer
architectures and quantum communication protocols \cite{Kimble2008,Huang2011,Fan2015}.  By its definition, the concept of cascading has been widely implemented in setups based upon the interconnecting of
quantum state sources and memories \cite{Lvovsky2009,Bussieres2013}.  However, protocols or operations demanding another degree of cascading, that is
setups that interconnect sources and multiple devices (i.e. memories) in a sequential manner have been primarily unexplored.

\begin{figure}[htb]
\includegraphics[width=1\columnwidth]{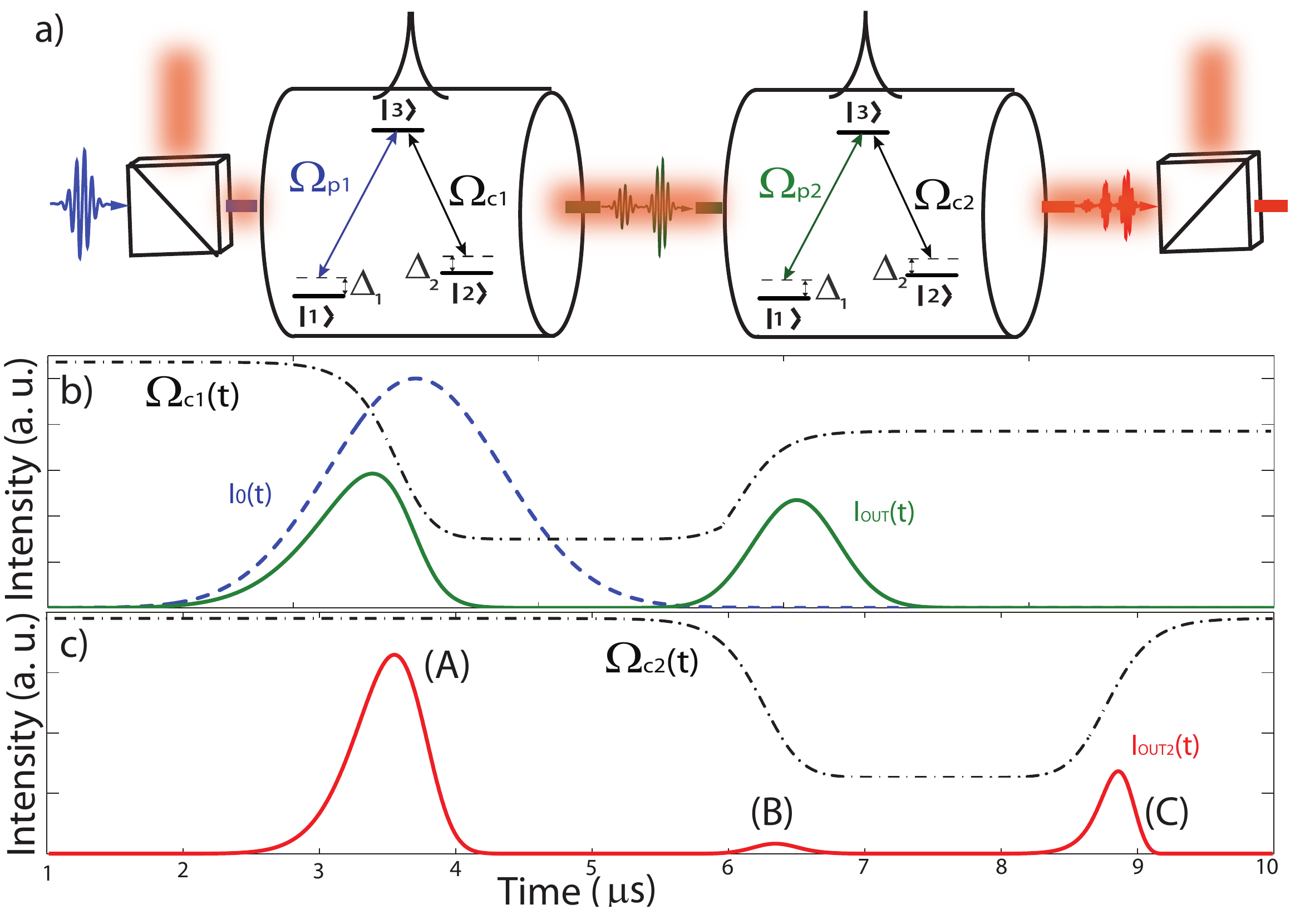} \protect\protect\caption{(a) Concept of the cascaded storage with two light-matter interfaces. (b) \textbf{First storage}: Input pulse (blue dotted line), Control field 1 time sequence (black dotted line) and retrieved light signal (solid green line) as obtained by simulation. (c) \textbf{Second storage}: Control field 2 time sequence (dotted black lines) and cascaded retrieved signal (solid red line).}
\end{figure}

Of the existing multi-device protocols, many will be reliant on operational quantum memories \cite{Monroe2014}, and furthermore on the functionality to cascade these devices, i.e. to have quantum
memories that efficiently interface with the output of a preceding memory. More specifically, cascading of quantum memories are necessary for certain one-way quantum computing schemes via clusters states with memory-assisted feed-forward operations \cite{Xu2012}, the implementation of conditional CZ gates utilizing quantum optical memories connected in series \cite{Campbell2014} and generating multi-mode quantum states by cascading multiple four-wave mixing processes in atomic ensembles \cite{Cai2014}.
The foundation of these implementations will require cascaded storage and retrieval schemes that exhibit both primary and secondary, high fidelity (with
respect to the original input) quantum memory operations. This demands the output of the first operation to
be a suitable input for a second memory operation.  Built on recent successes \cite{Novikova2011,Eisaman2005,Sprague2014,Hosseini2011_2}, we consider room temperature atomic vapor memories as the elements that comprise a series of cascadable devices that could form the foundation of a quantum network.  Room temperature systems are a promising direction,
as they can offer a relatively inexpensive experimental overhead while
also having strong light matter interaction at the single photon level
\cite{Michelberger2014,England2014,Kupchak2015}.

\begin{figure*}
\includegraphics[width=2\columnwidth]{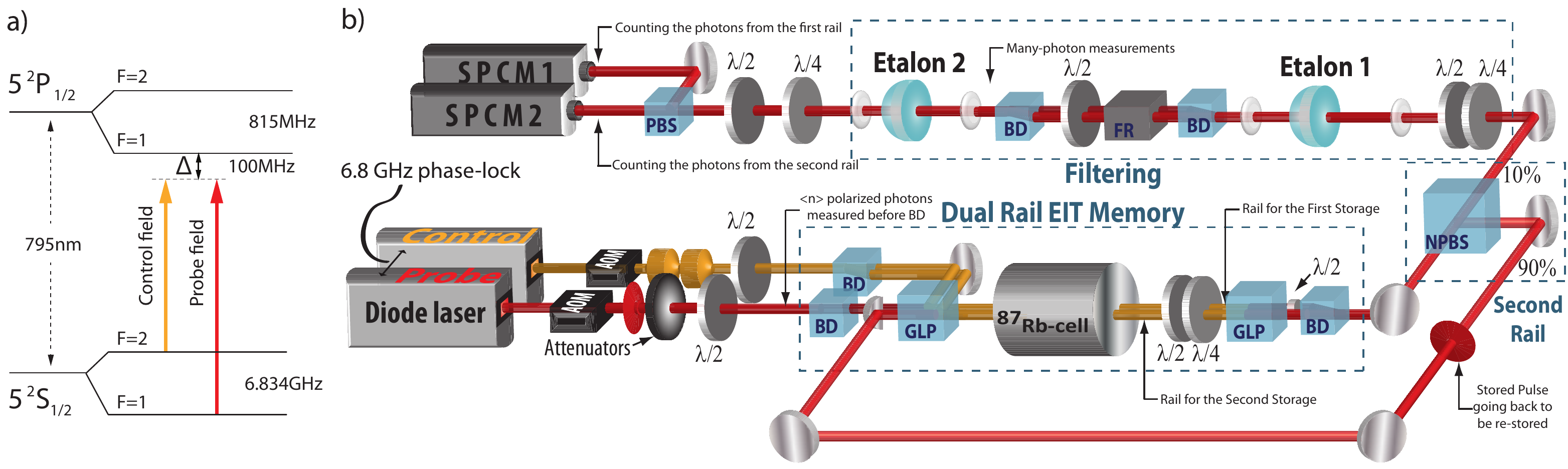}\protect\protect\caption{(a) Atomic level scheme and
EIT configuration used in both memories. (b) Experimental setup for successive storage
of pulses at the few-photon level, including the stages of control-filtering.
AOM: Acousto-optical modulators; BD: Beam displacers; GLP: Glan-Laser-Polarizer;
FR: Faraday rotator; SPCM: Single-Photon-Counting-Module; L: Lens;
M: Mirror; NPBS: Non-Polarizing Beam Splitter. Probe: red beam paths;
control: yellow beam paths. The NPBS transmits 10\% of the first stored pulse through the filtering system to be characterized and sends 90\% back
to the second rail for a successive storage.}
\end{figure*}


Here we present the cascaded storage of weak optical pulses containing
a few photons on average in two, room-temperature quantum light-matter
interfaces. Our implementation is based upon independently controlled, warm vapor ensembles prepared in the conditions of electromagnetically induced transparency (EIT) for the first and second storage procedures.  We determine under what conditions our setup produces an optimal signal-to-background ratio (SBR) for a cascaded retrieved light signal.  It has been shown that the overall fidelity of the cascaded operation will be closely tied to this SBR metric \cite{Kupchak2015}.

In order to utilize the technique of EIT for optical storage, we require atomic systems exhibiting a $\Lambda$-energy level scheme described by two ground states that resonantly couple to a common excited state.
In a cascaded optical storage procedure, the $\Lambda$-level scheme of the first atomic ensemble can be characterized
by the interaction with two laser fields, $\Omega_{p1}$ (probe) and
$\Omega_{c1}$ (control), with one-photon detunings $\Delta_{1}$
and $\Delta_{2}$ respectively (see Fig. 1). The output of this system will need to function as
an input to the second ensemble also exhibiting a $\Lambda$-type atomic level structure, and
characterized by the interaction with two laser fields, $\Omega_{p2}$
(retrieved probe coming from system 1) and $\Omega_{c2}$ (control).
We assume the detunings of both systems to be identical.

Optical storage in these $\Lambda$-systems can
be understood using the Hamiltonian which describes
the atom-field coupling in a rotating frame.  This is given by:
\begin{equation}
\hat{H}=\Delta_{1}\hat{\sigma}_{33}+(\Delta_{1}-\Delta_{2})\hat{\sigma}_{22}+\Omega_{p1}{E}_{p1}(z,t)\hat{\sigma}_{31}+\Omega_{c1}(t)\hat{\sigma}_{32}+h.c.
\end{equation}
where $\hat{\sigma}_{ij}=|i\rangle\langle j|,i,j=1,2,3$ are the atomic
raising and lowering operators for $i\neq j$, and the atomic energy-level
population operators for $i=j$ and ${E}_{p1}(z,t)$ is the normalized electric field amplitude of the probe. The dynamics of the first storage event
can be obtained numerically by solving the master equation for the
atom-light system density operator together with the Maxwell-Bloch
equation that contains the impact of the atomic polarization on the
electromagnetic field for a finite-length atomic sample
\begin{equation}
\dot{\hat{\rho}}=-i[\hat{H},\hat{\rho}]+\sum_{m=1,2}\Gamma_{3m}(2\hat{\sigma}_{m3}\rho\hat{\sigma}_{3m}-\hat{\sigma}_{33}\hat{\rho}-\hat{\rho}\hat{\sigma}_{33})
\end{equation}
\begin{equation}
\partial_{z}{E}_{p1}(z,t)=i\frac{\Omega_{p1}N}{c}\langle\hat{\sigma}_{31}(z,t)\rangle.
\end{equation}

Here $\Gamma_{31}$ and $\Gamma_{32}$ are the decay
rates of the excited level $|3\rangle$ to the ground states $|1\rangle$
and $|2\rangle$ respectively, $c$ is the speed of light in vacuum
and $N$ the number of atoms participating in the ensemble. Using initial conditions of
 $\Omega_{C1}(t)$ and $E_{p}(0,t)=E_{o}(t)$
(the original probe pulse shape) allows us to solve this set of equations
and calculate the expected retrieved pulse shape $E_{OUT}(t)$
= $E_{p1}(L,t)$, where $L$ denotes the length of the atomic ensemble.
Once we know $E_{OUT}(t)$, we can propagate this result to serve as the input
of the second $\Lambda$-system and similarly calculate the result of the cascaded storage procedure $E_{OUT2}(t)$.

In Figure 1b, we plot the simulation of the first storage and retrieval event using
$\Omega_{c1}(t)$ (black dotted line in Fig. 1b) and $E_{p}(0,t)=E_{o}(t)$
(blue dotted line in Fig. 1b) as the control and probe inputs respectively. The results for $E_{OUT}(t)$ = $E_{p1}(L,t)$
(Fig. 1b, in intensity, solid green light) is characteristic of an imperfect storage and retrieval signal, containing a portion of the input pulse that is not stored and transmitted
(left peak) followed by the retrieved signal due to the storage (right peak).

Figure 1c shows the simulation of the second storage and retrieval event
but now using $\Omega_{c2}(t)$ (dotted
black line in Fig. 1c, notice the time delay with respect to $\Omega_{c1}(t)$
to account for the first storage) and $E_{OUT}(t)$ as the control
and probe field inputs respectively.  The resultant cascaded stored signal contains three distinct peaks (Fig. 1c in intensity, solid red line), an initial probe leakage from the first storage procedure (A, leftmost peak),
a second small leakage from the second procedure (B, middle peak)
and a third peak whose timing matches that of when the second control field is switch on again (C, right most peak). This final peak corresponds
to a portion of the probe field that has been sequentially stored and retrieved
in two independent light matter interfaces and is the focus of this
letter. Note that in our simulations we have used $\Gamma_{31}=\Gamma_{32}= 3.0\pi*10^6 s^{-1}, N \sim 1^{10}$ atoms and $L=7 cm$.

In order to implement the aforementioned cascaded optical storage procedure experimentally, we
employed two external-cavity diode lasers as light sources, phase-locked
at 6.8 GHz to resonantly couple a $\Lambda$-configuration composed
of two hyperfine ground states sharing a common excited state. The
probe field frequency is stabilized to the $5S_{1/2}F=1$ $\rightarrow$
$5P_{1/2}F'=1$ transition at a wavelength of 795 nm (red detuning
$\Delta$=100 MHz) while the control field interacts with the $5S_{1/2}F=2$
$\rightarrow$ $5P_{1/2}F'=1$ transition (Fig. 2a).

Our optical setup is adapted from our prior dual-rail memory experiment
for polarization qubits \cite{Kupchak2015}, where each rail now
serves as a distinct optical memory contained in a single vapor cell. The temporal shaping of the probe
and the control field of each rail are independently controlled with
acousto-optical modulators and driven by arbitrary signal generators
for amplitude modulation. A polarization beam displacer is used to
create a dual-rail set-up for the control field where each rail is mode matched
to the respective probe via a Glan-laser polarizer (see Fig. 2b). An initial 100 $\mu$W-peak input
pulse of 1 $\mu$s duration  is fixed to horizontal polarization and sent through the first rail (solid blue line in Fig. 3). Using one
of the control fields, the probe pulse is stored in a room temperature
cell containing isotopically pure Rb 87 vapor for a duration of 1
$\mu$s using EIT for storage and retrieval. For the first storage, we apply a temporal modulation to the control field used for retrieval, which allows tuning of the instantaneous group
velocity of the retrieved excitation and consequently the tailoring
of its temporal shape. We engineer the control field amplitude to provide a retrieved pulse (from the atomic ensemble in rail
1) with a near Gaussian temporal profile (see 2nd peak of solid green
line in Fig. 3) to yield an efficiency ($\eta_{1}$) of $\sim$ 12\%.  Notice
that because the length of our vapor cell does not accommodate the
full length of the original input pulse, we have a leakage as predicted by our simulations (see
1st peak of the solid green line in Fig. 3).

The retrieved pulse is transmitted through a polarizer for filtering followed
by a beam displacer for recombination to a single beam path.
After this step, a 90/10 beam splitter is used to send the majority
of the retrieved photons back to the front of the vapor cell (see Fig. 2)
where a pick-off mirror sends the signal through the second rail. The timing of control field 2 is matched
to the retrieval of the first memory for the second storage
sequence. After the second beam displacer, the signals from the first and second-rail are
matched to the same beam path, albeit with orthogonal polarizations which permits independent measurement of both storage experiments. The signals transmitted through the 90/10 beam splitter continue through
a temperature-controlled etalon and a polarization independent Faraday-isolator to remove the remnants of the control field.  At this point, the classical-level signals are detected independently in a photo-detector (not shown in Fig. 2). The signals from the second rail (blue and red in Fig. 3) are 3.9 times
smaller than those from the first (green in Fig. 3) due to propagation losses and mismatched etalon coupling efficiencies.
As shown in Figure 3 (red line), the resultant cascaded
stored signal has three peaks as was predicted by our simulations
and is, to our knowledge, the first time that such cascaded storage process
has been demonstrated.

\begin{figure}[htb]
\includegraphics[width=1\columnwidth]{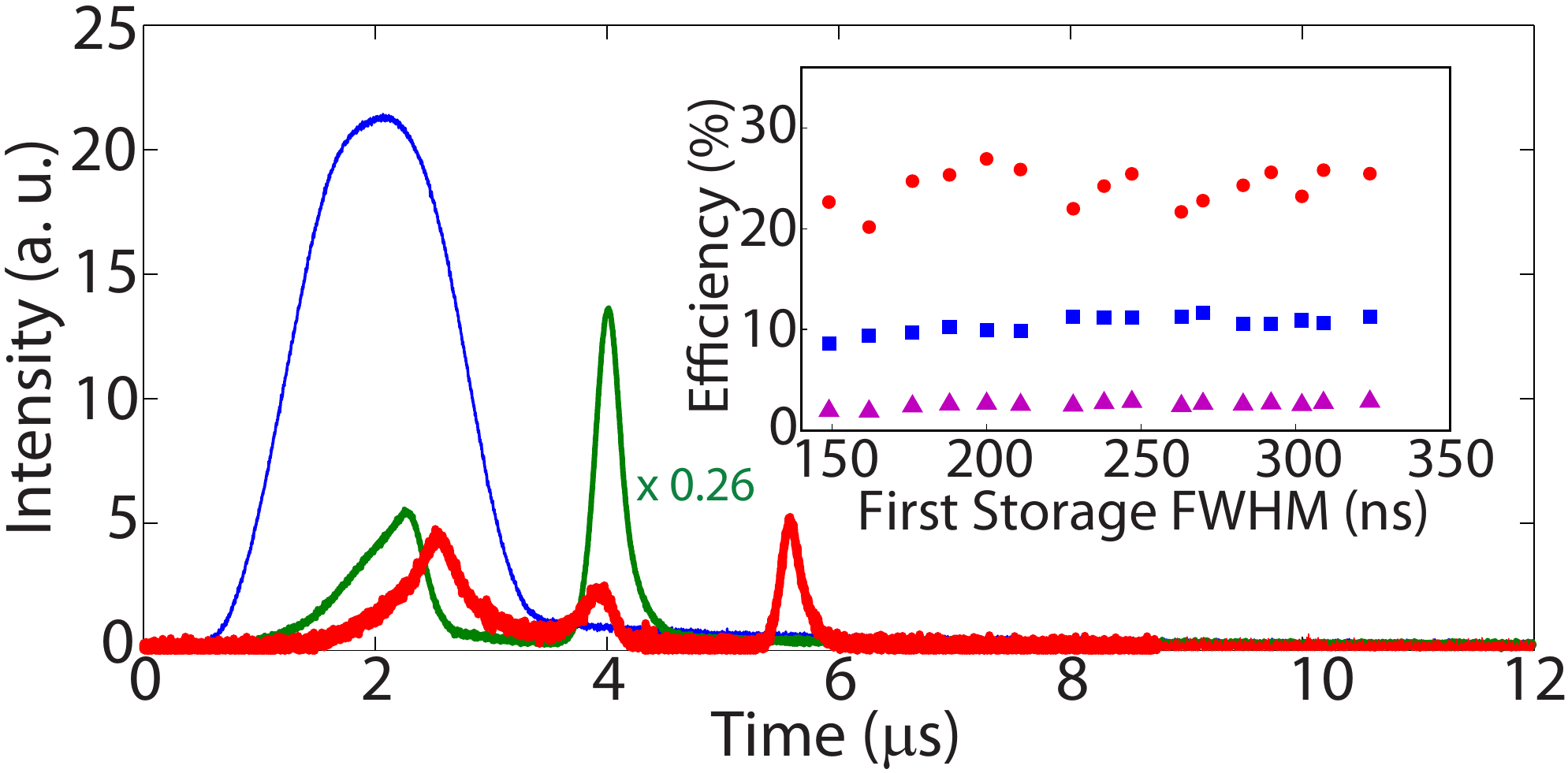}
\caption{Successive storage of classical pulses. Blue:
input pulse; Green: rail 1 single storage signal; Red: rail 2 cascaded storage signal.
The green line is scaled by a factor 0.26 to account for different propagation losses after the first and second rails.
Inset: dependency of storage efficiencies on the full-with half maximum of the Gaussian shaped retrieved field for the first storage.
Blue: efficiency of the first storage $\eta_{1}$; Red: efficiency of the second storage $\eta_{2}$; Purple:
overall efficiency of cascaded storage procedures $\eta_{T}$.}
\end{figure}

To maximize the efficiency of the cascaded storage ($\eta_{T}$),
we modify the duration of the control field used for the first retrieval
which also affects the temporal length of the retrieved probe field.
This has a significant effect on $\eta_{T}$, as the optimal bandwidth
of the retrieved pulse resembles the EIT bandwidth exhibited by
the vapor cell. A total storage efficiency of $\sim$ 3\% is obtained
when the duration of the control field for the first retrieval is
300ns. The efficiency of the second storage event is independently verified to be $\eta_{2} \sim 25\%$.

Now that we have demonstrated our ability to successively store classical light
pulses, we turn our attention to operating our system at the few-photon
level. Specifically, we are interested in benchmarking the behaviour
of the complete optical storage network and determine the
parameters needed to obtain a cascaded retrieved signal (at the end of the
network) that is at the same level of the background produced by the
experiment i.e. signal-to-background ratio (SBR) of 1.

To do so, we probe our system with coherent state pulses at
the few photon level. A trace of the input state is shown in Figure
4 (solid green line, from 1 to 2 $\mu$s) for an input mean photon
number $\sim$ 8. However, in order to sufficiently extinguish
the large number of photons coming from the control fields, we
add a second filtering etalon to the setup of the previous measurement.
Overall, the complete filtering setup achieves 154 dB of control field suppression, including
the 90/10 beam splitter, while yielding a total probe field
transmissions of 0.39\% and 0.22\% for the first and second rails respectively, to generate an effective, control/probe suppression ratio
of about 130 dB. As discussed before, the setup permits measurement of the storage in the first rail (see SPCM 1 in Fig. 2) or the cascaded storage
from the second rail (see SPCM 2 in Figure 2).

\begin{figure}[htb]
\includegraphics[width=1\columnwidth]{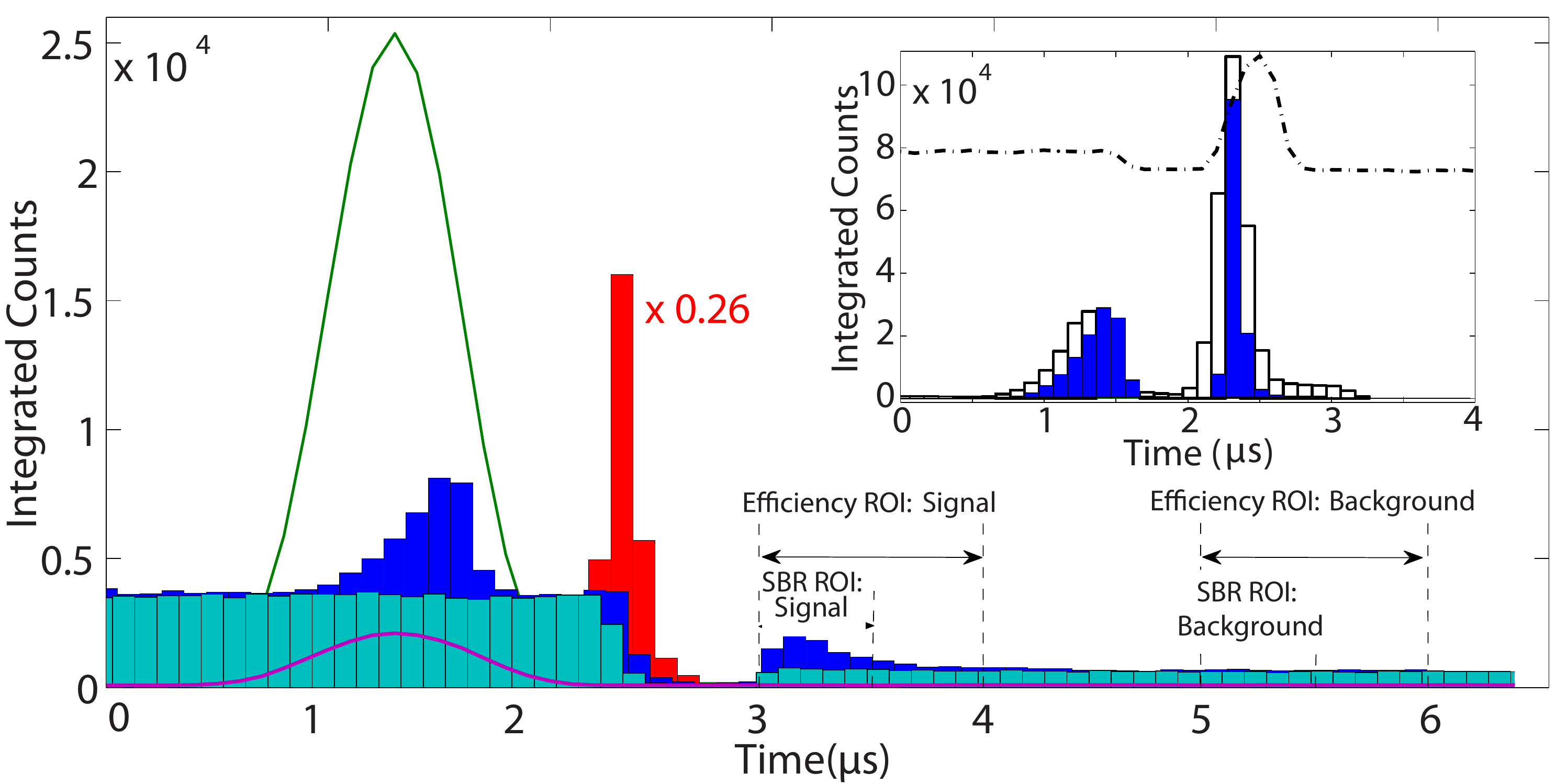}
\protect\protect\caption{Cascaded optical storage for input pulses containing 8 photons per pulse on average. Green: input pulse;
Purple: absorbed pulse (after passage through both rails); Red: retrieved pulse after first storage (scaled by a factor 0.26); Blue:
retrieved signal after cascaded storage procedure (with background); Light blue: background-only measurement.
Inset: The effect of reshaping the control field for the optical retrieval in the first rail.  Storage using TTL driven control field (blue bars)
and storage counts obtained with a temporally modulated control field
(green bars). The temporal shape of the control field is indicated by the dotted-black line.}
\end{figure}

To determine the total storage efficiency ($\eta_{1}$) in the first rail,
we integrate the number of counts over the region of interest (ROI)
corresponding to the retrieved pulse (from 2 to 2.5 \textmu s in the
inset of Fig. 4) and subtract the number of counts from a signal-free
measurement of the background over the same ROI. The magnified background shape of the control field is included
(dashed black line in the inset of Fig. 4) as a guide to the eye.
The storage efficiency is then calculated by comparing this difference in counts to the total counts
of the probe pulse transmitted (in rail 1) through the filtering system without atomic
interaction. The signal to background ratio is obtained in a similar
fashion using the counts integrated over the ROI in the storage histogram
(signal+background) and the number of counts over a signal-free region
in the same histogram (background). The SBR is then calculated as
{[}(signal+background)-(background){]}/(background).

As shown in the inset of Figure 4, there is a considerable effect
on $\eta_{1}$ by using a temporal shaping of the control field for retrieval
(white bars), as compared to an experiment in which the control
field is driven with a TTL signal (blue bars).  In this instance, we find a maximum signal-to-background ratio of 13 and an
efficiency of 14.6\% (see red histogram in Fig. 3) from our input state.

The majority of the photons retrieved from the first
memory are sent to the second rail \emph{together with any photons from the first control field} without passing through the filtering setup.
We find that after propagation losses (including the routing beam
splitter and interconnecting losses from the first to second rail) of 53.4\%,
the mean photon number of the probe field at the input of the second
memory is 0.6 photons compared to $\sim$ 10$^{8}$ photons per pulse
from the background. The probe photons are re-stored and then retrieved
using the second control field (see dark blue histogram in Fig. 4). For comparison,
we also show the counts recorded when the input has been blocked (see
light blue bars in Fig. 4).

The cascaded storage signal has a SBR of 1.2 for our input
state. Using a similar procedure to the
one described previously, we measure the overall
efficiency of the cascaded storage (using a ROI in the interval from
3 to 4 $\mu$s in Figure 3) to be $\eta_{T}=3.2\%$. The efficiency of the second memory $\eta_{2}$ was found to
be 21.7\%.

\begin{figure}[htb]
\includegraphics[width=1\columnwidth]{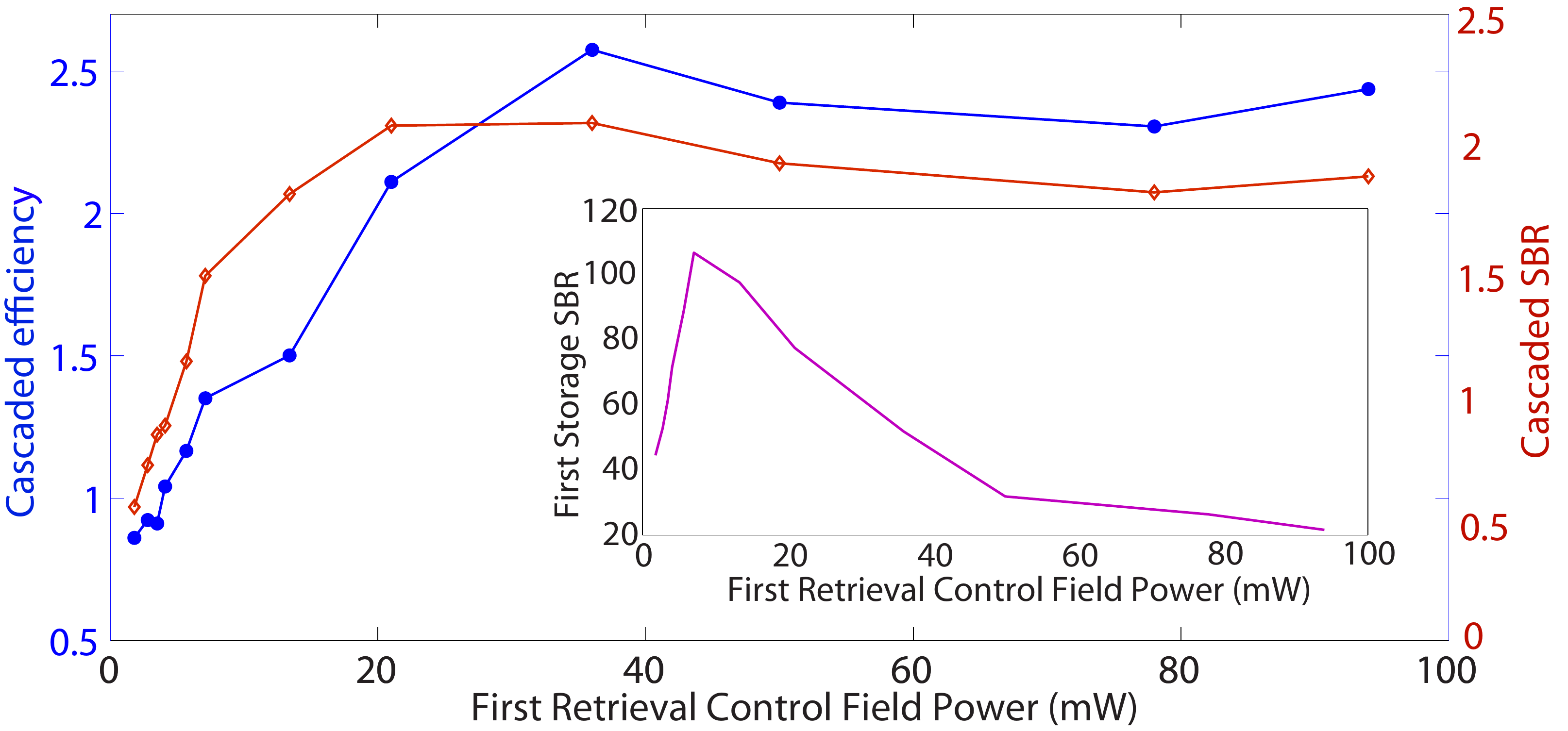}
\protect\protect\caption{Total cascaded storage efficiency $\eta_{T}$ (blue line with circles) and cascaded SBR (red line with diamonds) vs. optical power for first retrieval. Inset: SBR of first storage event vs. optical power for first retrieval.}
\end{figure}

Finally we turn our attention to the noise characteristics of our
cascaded storage system. Specifically, we are interested in the influence
of the background noise photons generated from the first optical storage
event on the final cascaded storage signal (after memory 2). To do
so, we have measured the cascaded storage efficiency ($\eta_{T}$), the cascaded SBR and the SBR of the first storage event  vs. the control field power used for retrieval in the first rail.  In order to obtain a more precise
SBR and efficiency, we used input states containing an average of
18 photons per pulse and a TTL driven control field for the retrieval
(oppose to the reshaping method described). In this way, any noise
photons generated when the first control field is turned on (either
leakage or atomic-triggered background) can be measured at the
time of the cascaded storage signal.

We can see that the total SBR after the cascading
event (solid red line in Fig. 5) follows the behavior of the total
storage efficiency $\eta_{T}$ (solid blue line in Fig. 5). By comparison,
we can see that the SBR for the first storage event (see inset in
Fig. 5) does not follow the SBR of the cascaded procedure. This indicates
that the second EIT storage ensemble additionally serves as frequency
filter of the background noise generated from the first storage ensemble.
From a quantum engineering point of view this becomes a interesting
aspect, implying that filtering schemes may only be needed at the point of final measurement
readout after a cascaded set of operations. A future network of multiple devices that contains a built-in filtering
mechanism inherent to the nature of the system could indeed be beneficial over
a setup that requires additional filtering and clean-up
hardware and operations after each individual task. The availability
of such self-filtering systems would
lead to a decrease in both experimental overhead and overall loss in the end-point read-out signal
and becomes a major consideration when constructing networks of this type.

In summary, we have demonstrated the cascaded storage of few-photon
level pulses using two distinct room temperature ensembles contained
in the same vapor cell. Our results demonstrate that with our current room temperature technology
it is viable to interconnect two quantum light-matter
interfaces in a sequential manner, a key attribute of a quantum optical
network.

In our particular implementation it is not possible to use an original input
at the single-photon level due to the inherent propagation and interconnection losses required by using a dual rail system.
This shortcoming can be easily bridged by using fully independent memory systems (separate vapor cells) connected in series. Achieving this interconnection between quantum memories for input single photons carrying qubits could be a milestone towards building more sophisticated machines that interface even more quantum optical nodes. This in turn will pave the way for the creation of elementary one-way quantum information processors based on warm vapor ensembles.

This work was supported by the US-Navy Office of Naval Research, grant
number N00141410801 (instrumentation) and the National Science Foundation,
grant number PHY-1404398 (personnel and materials). C. K. acknowledges
financial support from the Natural Sciences and Engineering Research
Council of Canada.

\end{document}